\documentclass[pdflatex]{acta}
\usepackage{amssymb}
\usepackage{amsmath}
\usepackage{lscape}
\usepackage{rotating}

\newcommand{\degg}[0]{^{\circ}}

\SetPages{0}{0}

\SetVol{69}{2019}

\begin{document}

\begin{Titlepage}
\Title{Mapping the Northern Galactic Disk Warp with Classical Cepheids.}
\Author{
D.~M.~~S~k~o~w~r~o~n$^1$,
~~J.~~S~k~o~w~r~o~n$^1$,
~~P.~~M~r~{\'o}~z$^{1,2}$,
~~A.~~U~d~a~l~s~k~i$^1$,
~~P.~~P~i~e~t~r~u~k~o~w~i~c~z$^1$,
~~I.~~S~o~s~z~y~{\'n}~s~k~i$^1$,
~~M.~K.~~S~z~y~m~a~{\'n}~s~k~i$^1$,
~~R.~~P~o~l~e~s~k~i$^{1,3}$,
~~S.~~K~o~z~{\l}~o~w~s~k~i$^1$,
~~K.~~U~l~a~c~z~y~k$^{1,4}$,
~~K.~~R~y~b~i~c~k~i$^1$,
~~P.~~I~w~a~n~e~k$^1$,
~~M.~~W~r~o~n~a$^1$
~~and~~M.~~G~r~o~m~a~d~z~k~i$^1$}
{$^1$ Astronomical Observatory, University of Warsaw, Al.~Ujazdowskie~4, 00-478~Warszawa,
Poland\\
$^2$ Division of Physics, Mathematics, and Astronomy, California Institute
of Technology, Pasadena, CA~91125,~USA\\
$^3$ Department of Astronomy, Ohio State University, 140 W. 18th Ave.,
Columbus, OH~43210,~USA\\
$^4$ Department of Physics, University of Warwick, Gibbet Hill Road,
Coventry, CV4~7AL,~UK\\
e-mail: dszczyg@astrouw.edu.pl
}

\end{Titlepage}

\Abstract{
  We present an updated three dimensional map of the Milky Way based
  on a sample of 2431 classical Cepheid variable stars, supplemented
  with about 200 newly detected classical Cepheids from the OGLE survey.
  The new objects were discovered as a result of a dedicated observing
  campaign of the $\sim$280 square degree extension of the OGLE footprint
  of the Galactic disk during 2018--2019 observing seasons. These regions
  cover the main part of the northern Galactic warp that has been deficient
  in Cepheids  so far.
  We use direct distances to the sample of over 2390 classical Cepheids
  to model the distribution of the young stellar population in the Milky
  Way and  recalculate the parameters of the Galactic disk warp. Our data
  show that its northern part is very prominent and its amplitude is
  $\sim$10\% larger  than that of the southern part.
  By combining \textit{Gaia} astrometric data with the Galactic rotation
  curve and distances to Cepheids from our sample, we construct a map of the
  vertical component of the velocity vector for all Cepheids in the Milky Way
  disk. We find large-scale vertical motions with amplitudes of 10--20 km/s,
  such that Cepheids located in the northern warp exhibit large positive
  vertical velocity (toward the north Galactic pole), whereas those in the
  southern warp -- negative vertical velocity (toward the south Galactic
  pole).
}

\section{Introduction}

The warping of the Milky Way disk has been detected with multiple tracers:
21~cm observations of neutral Hydrogen (\eg Burke 1957; Westerhout 1957;
Nakanishi \& Sofue 2003; Levine \etal 2006), distribution of dust
(\eg Marshall \etal 2006) and stars (\eg Drimmel \& Spergel 2001;
Reyl\'e \etal 2009; Am\^ores \etal 2017), or stellar kinematics
(\eg Smart \etal 1998; Poggio \etal 2018; Romero-G\'omez \etal 2019).
However, the distances of these tracers are model-dependent.
Only the studies of Galactic classical Cepheids have mapped the warp
with directly measured distances to individual stars (Berdnikov 1987,
Skowron \etal 2019, Chen \etal 2019, D\'ek\'any \etal 2019).

We know that warped stellar disks are common -- over 50\% of spiral
galaxies experience some degree of warping (Sanchez-Saavedra \etal 1990),
but the mechanism standing behind this phenomenon is unknown.
The origin of the Milky Way warp is still debated and possible explanations
can be divided in two broad classes. One possibility is that the warp formed
as a result of gravitational interactions, for example, with satellite
galaxies or a mis-aligned dark matter halo. Other models propose
non-gravitational mechanisms such as accretion of intergalactic gas or
interactions with intergalactic magnetic fields as a possible explanation
of warping. See L\'opez-Corredoira (2019) and references therein.
In the former models, gas as well as old and young stellar populations should
be warped in a similar pattern. Alternatively, one should observe an age
dependency of the Galactic warp. It is therefore important to identify
individual stars of different age in the warp and measure their precise
distances.

The most complete and pure sample of classical Cepheids in the Milky
Way from the Optical Gravitational Lensing Experiment (OGLE) long-term
sky survey (Udalski \etal 2018), supplemented with Cepheids from other
surveys has been recently used by Skowron \etal (2019) to construct the
most detailed 3-D map of the Galactic disk in the young stellar population,
extending out to the edge of the Galaxy. The map allowed constraining the
shape of the warp and the location of the line of nodes that do not match
models based on other stellar populations (Romero-G\'{o}mez \etal 2019,
Am\^ores \etal 2017).

Cepheids used by Skowron \etal (2019) covered the southern part of the
Galactic warp (bent toward the south Galactic pole), in the Galactic
longitude range $180\degg-360\degg$. However, the northern side of the Milky
Way warp (bent toward the north Galactic pole and located in the first Galactic
quadrant, \ie $0\degg<l<90\degg$), has been poorly constrained due to low
number of classical Cepheids in that area. This is due to extremely high
extinction close to the Galactic center and the $l<40\degg$ limit of the
OGLE Collection of Galactic Cepheids. However, the constraints of the OGLE
observing site (Las Campanas Observatory, Chile) and the OGLE telescope
limitations allow obtaining reasonable photometry of Milky Way disk stars
up to $l<60\degg$.  Therefore it would be possible to extend the Cepheid
sample with the first quadrant Cepheids by expanding the range of the OGLE
footprint in the Galactic disk.

Knowing that the regions of high extinction in the first quadrant of the
Galaxy may be too opaque for optical observations and to estimate the
detectability of classical Cepheids in the OGLE extended survey, we
carried out simulations in which we generated a sample of 100\,000
artificial Cepheids from a 3-D distribution based on the disk model
presented by Skowron \etal (2019). We assigned them absolute $I$-band
magnitudes of $-3.2$ and $-4.8$ (corresponding to the pulsation periods
of 3 and 10 days, respectively; Gieren \etal 1998). We also used the 3-D
extinction maps (Bovy \etal 2016) to estimate the $I$-band extinction
toward each source. We treated a Cepheid as detected if its apparent
magnitude is in the range $10.75<I<18.0$.  Fig.~1 shows the detection
probability for $M_I=-3.2$ mag (left) and $M_I=-4.8$ mag (right). The dashed
line marks the original range of the OGLE survey.  We see that extending the
survey limit in the first Galactic quadrant, where the warping is the most
significant, would ensure new detections.

\vspace{0.1cm}
\begin{figure*}[ht]
\centerline{\includegraphics[width=13cm]{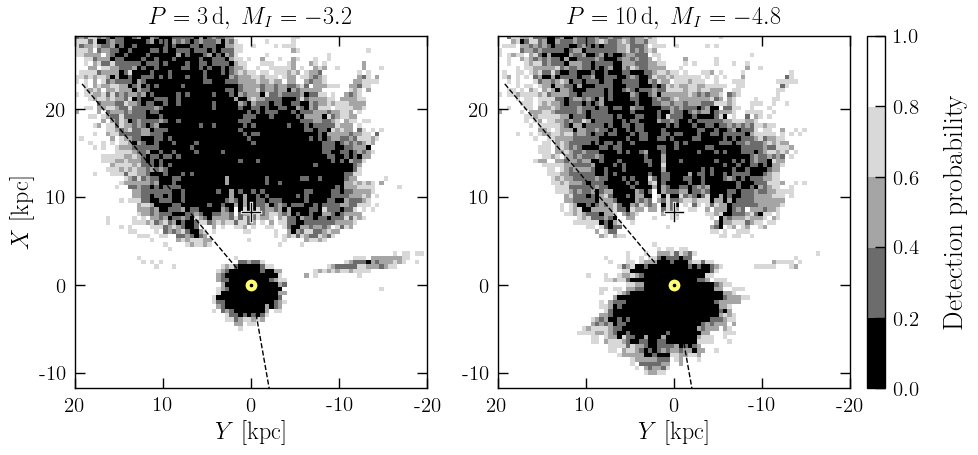}}
\FigCap{The detection probability of classical Cepheids by the OGLE
  telescope for absolute $I$-band magnitudes of $-3.2$ (left) and
  $-4.8$ (right), corresponding to pulsation periods of 3 and 10 days,
  respectively.  The Sun is marked with a yellow dot while the
  Galactic center with a cross.  The dashed line marks the original
  range of the OGLE survey ($l<40\degg$ and $l>190\degg$).  }
\vspace{0.3cm}
\label{fig:simulation}
\end{figure*}

\section{Data Preparation and Analysis}

\subsection{Observational Data}

Motivated by the outcome of the detectability simulations, the OGLE project
has continued its Galaxy Variability Survey (GVS, Udalski \etal 2015) aiming
at supplementing the OGLE Collection of Galactic Cepheids with new objects and
detecting other types of variable objects in the Galactic longitude range
$20\degg<l<60\degg$. During the 2018 and 2019 observing seasons 201 new fields
were monitored in the extended area. The number of new epochs reached 60--120
per field, depending on the location, making 122 fields suitable for the
regular OGLE variability analysis. Some of these fields have additional
observations from seasons 2015--2017.  The process of data
reduction and classical Cepheid classification in these fields was the same
as in Udalski \etal (2018).

In the case of the remaining 79 fields the number of collected epochs was much
lower -- reaching 10--20 observations per field. This is not sufficient for
conducting a regular OGLE variability search. Nevertheless, we attempted to
extract objects which reveal light curves with an unambiguous classical Cepheid
shape from the sample of clearly variable objects. The main difficulty
was the estimation of the correct pulsation period based on a very small number
of epochs. To make sure that our selection of classical Cepheids from these
79 fields is reliable we verified their photometry in publicly available data
of the All Sky Automated Survey (ASAS; Pojma{\'n}ski 2002) and the Zwicky
Transient Facility (ZTF; Masci \etal 2018), confirming most of the new
candidates.  Our search in these fields is in principle less complete than in
the remaining parts of the sky covered by OGLE. Nevertheless, the purity of
the sample should be retained.

We identified 223 classical Cepheids in the extended footprint of the OGLE
GVS, of which 23  were already listed in the sample of Skowron \etal (2019).
We cross-checked the list with other catalogs and found two on the {\em Gaia}
Cepheid list (Holl \etal 2018), four in the WISE variable stars
catalog (Chen \etal 2018) and one on the list of Chen \etal (2019).
Fig.~2 presents the on-sky locations of classical Cepheids from the Skowron
\etal (2019) sample (yellow and blue dots), supplemented with new detections
(magenta dots), while Table~1 provides the parameters for a subset of the new
list. The on-line version of the OGLE Collection of Galactic Cepheids
(Udalski \etal 2018) has already been supplemented with the new objects.

\vspace{0.1cm}
\begin{landscape}
\begin{figure*}[ht]
\centerline{\includegraphics[width=20cm]{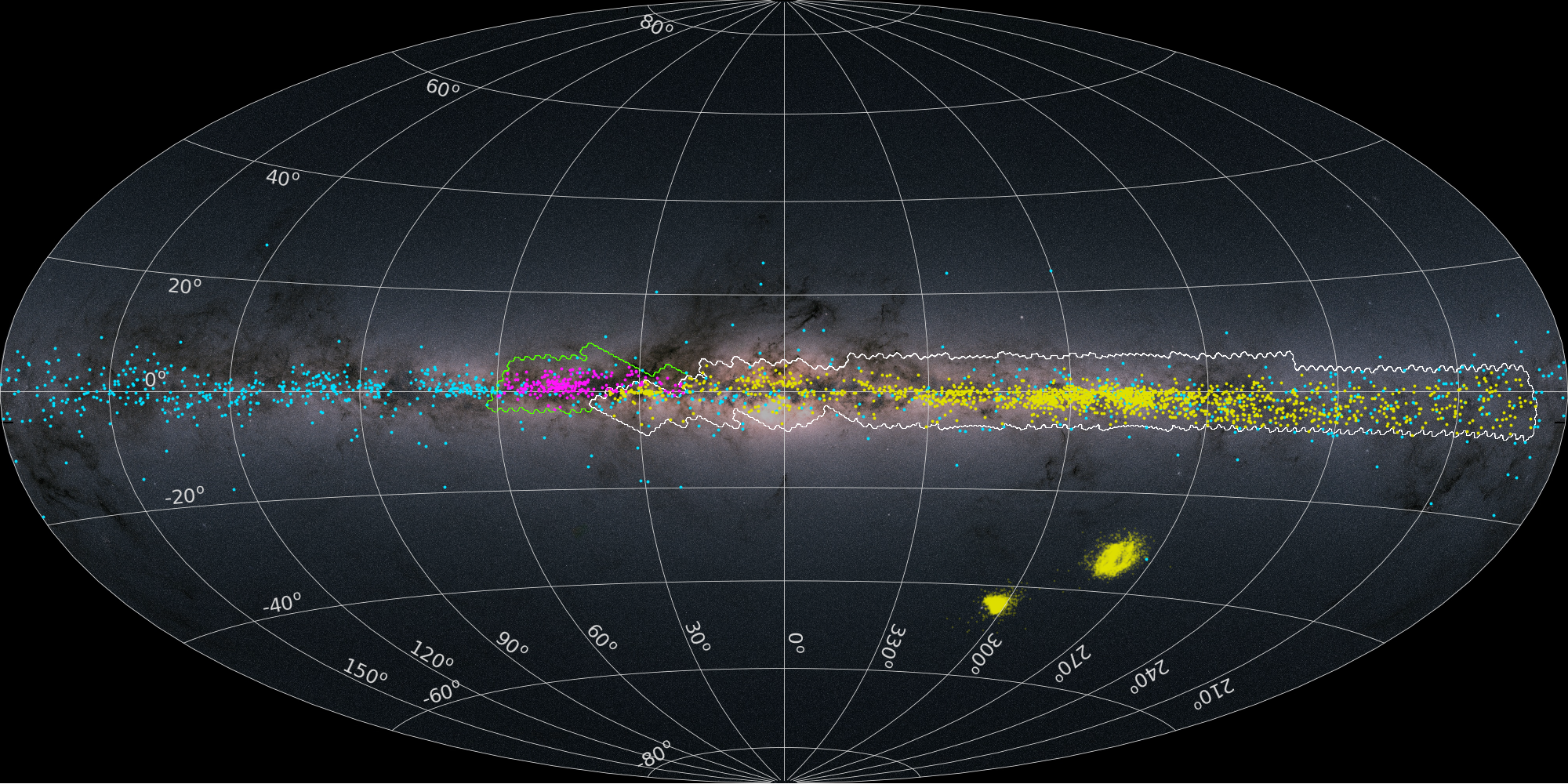}}
\FigCap{The on-sky view of the Milky Way in the Galactic coordinates,
  on top of the Milky Way image (Credit: ESA/Gaia/DPAC, CC BY-SA 3.0
  IGO\footnote{https://creativecommons.org/licenses/by-sa/3.0/igo/})
  with all known Classical Cepheids from the OGLE collection (yellow dots)
  and from other surveys (cyan dots).  Newly detected objects are marked
  with magenta dots. The total number of objects is 2632. The white
  contour marks the past OGLE survey footprint of the Galactic plane
  while the green one -- the extension analyzed here\footnote{This image
  can be distributed under the CC BY-SA 3.0 licence}.}
\vspace{0.3cm}
\label{fig:side}
\end{figure*}
\end{landscape}

\begin{sidewaystable}
\tiny
\begin{tabular}{|ll|rrrr|rrr|rrrr|rrrr|r|}
\hline
Source              & Cepheid ID            &       $l$    &        $b$   &    $d$  &$\sigma_d$&   Age  &      Period    &F/1O&   I1     &   I2     &   I3     &   I4     &   W1     &   W2     &   W3     &   W4     &${\rm A}_{\rm Ks}$\\
                    &                       &      [deg]   &       [deg]  &    [pc]  &   [pc]  &  [Myr] &       [d]      &    &  [mag]   &  [mag]   &  [mag]   &  [mag]   &  [mag]   &  [mag]   &  [mag]   &  [mag]   &  [mag]\\
\hline
OGLE-IV             & OGLE-GD-CEP-1367        & $  20.93065$ & $  -0.36615$ & $  5360$ & $  120$ & $  48$ & $  17.4641618$ & F  & $ 7.542$ & $ 7.515$ & $ 7.408$ & $ 7.384$ & $ 7.666$ & $ 7.588$ & $ 7.546$ & $ 0.000$ & $0.821$ \\
OGLE-IV             & OGLE-GD-CEP-1361        & $  21.36920$ & $   1.65681$ & $    -1$ & $    0$ & $ 103$ & $   4.2078655$ & 1O & $-1.000$ & $-1.000$ & $-1.000$ & $-1.000$ & $-1.000$ & $-1.000$ & $-1.000$ & $-1.000$ & $0.000$ \\
OGLE-IV             & OGLE-GD-CEP-1363        & $  21.62651$ & $   1.17220$ & $    -1$ & $    0$ & $ 174$ & $   2.5083880$ & F  & $-1.000$ & $-1.000$ & $-1.000$ & $-1.000$ & $-1.000$ & $-1.000$ & $-1.000$ & $-1.000$ & $0.000$ \\
OGLE-IV             & OGLE-GD-CEP-1366        & $  21.78451$ & $   0.08713$ & $  7634$ & $  305$ & $  33$ & $  37.1790749$ & F  & $ 7.457$ & $ 7.430$ & $ 7.226$ & $ 7.207$ & $ 7.815$ & $ 7.641$ & $ 7.187$ & $ 4.691$ & $1.416$ \\
OGLE-IV             & OGLE-GD-CEP-1364        & $  22.64066$ & $   1.22373$ & $ 16804$ & $ 1289$ & $ 126$ & $   4.3574838$ & F  & $-1.000$ & $-1.000$ & $-1.000$ & $-1.000$ & $11.957$ & $12.092$ & $10.056$ & $ 0.000$ & $0.938$ \\
OGLE-IV             & OGLE-GD-CEP-1377        & $  22.69182$ & $  -0.82917$ & $  4125$ & $  104$ & $  58$ & $  12.9505463$ & F  & $ 7.447$ & $ 7.366$ & $ 7.288$ & $ 7.282$ & $ 7.110$ & $ 7.222$ & $ 7.258$ & $ 6.579$ & $0.787$ \\
OGLE-IV             & OGLE-GD-CEP-1372        & $  22.96984$ & $   0.12494$ & $  3210$ & $   62$ & $  79$ & $   4.8895263$ & 1O & $ 7.525$ & $ 7.493$ & $ 7.413$ & $ 7.374$ & $ 7.485$ & $ 7.420$ & $ 7.262$ & $ 5.912$ & $0.420$ \\
ATLAS               & J279.0161-07.9499       & $  24.11004$ & $  -0.22832$ & $  3363$ & $  153$ & $  58$ & $  13.2058389$ & F  & $ 0.000$ & $ 6.809$ & $ 6.634$ & $ 6.588$ & $ 6.739$ & $ 6.871$ & $ 7.178$ & $ 0.000$ & $0.750$ \\
OGLE-IV             & OGLE-GD-CEP-1362        & $  24.22708$ & $   2.77176$ & $ 23338$ & $  880$ & $ 258$ & $   1.9732554$ & 1O & $13.111$ & $13.195$ & $12.226$ & $12.960$ & $13.201$ & $13.587$ & $ 0.000$ & $ 0.000$ & $0.460$ \\
OGLE-IV             & OGLE-GD-CEP-1378        & $  24.74936$ & $   0.13698$ & $  4661$ & $  126$ & $  58$ & $  12.8564317$ & F  & $ 7.818$ & $ 7.672$ & $ 7.509$ & $ 7.583$ & $ 7.794$ & $ 7.729$ & $ 6.457$ & $ 3.871$ & $1.025$ \\
OGLE-IV             & OGLE-GD-CEP-1369        & $  25.48430$ & $   1.58357$ & $ 10494$ & $  287$ & $ 139$ & $   2.5164794$ & F  & $11.763$ & $11.624$ & $11.564$ & $11.587$ & $11.739$ & $11.845$ & $ 0.000$ & $ 0.000$ & $0.741$ \\
OGLE-IV             & OGLE-GD-CEP-1368        & $  25.78035$ & $   1.91519$ & $ 20438$ & $ 1144$ & $ 202$ & $   2.6340611$ & F  & $13.094$ & $12.944$ & $ 0.000$ & $ 0.000$ & $-1.000$ & $-1.000$ & $-1.000$ & $-1.000$ & $0.619$ \\
OGLE-IV             & OGLE-GD-CEP-1365        & $  26.26834$ & $   2.78097$ & $    -1$ & $    0$ & $ 212$ & $   1.7844657$ & F  & $-1.000$ & $-1.000$ & $-1.000$ & $-1.000$ & $-1.000$ & $-1.000$ & $-1.000$ & $-1.000$ & $0.000$ \\
OGLE-IV             & OGLE-GD-CEP-1374        & $  28.77625$ & $   2.86816$ & $    -1$ & $    0$ & $ 172$ & $   2.5380653$ & F  & $-1.000$ & $-1.000$ & $-1.000$ & $-1.000$ & $-1.000$ & $-1.000$ & $-1.000$ & $-1.000$ & $0.000$ \\
GCVS                & DV\_\_\_\_Ser           & $  29.52992$ & $   1.40697$ & $  6480$ & $  166$ & $  42$ & $  23.1308428$ & F  & $ 7.344$ & $ 7.335$ & $ 7.334$ & $ 7.249$ & $ 7.140$ & $ 7.194$ & $ 7.199$ & $ 0.000$ & $0.422$ \\
OGLE-IV             & OGLE-GD-CEP-1384        & $  29.59993$ & $   1.48647$ & $ 14376$ & $  452$ & $ 114$ & $   4.6961076$ & F  & $11.414$ & $11.357$ & $11.296$ & $11.433$ & $-1.000$ & $-1.000$ & $-1.000$ & $-1.000$ & $0.594$ \\
OGLE-IV             & OGLE-GD-CEP-1382        & $  29.63869$ & $   1.97264$ & $ 19625$ & $  821$ & $ 132$ & $   5.0145837$ & F  & $12.166$ & $12.053$ & $11.853$ & $11.817$ & $-1.000$ & $-1.000$ & $-1.000$ & $-1.000$ & $0.685$ \\
OGLE-IV             & OGLE-GD-CEP-1386        & $  29.95373$ & $   1.33633$ & $ 20479$ & $  521$ & $  90$ & $   6.5061474$ & 1O & $11.386$ & $11.276$ & $11.192$ & $11.263$ & $-1.000$ & $-1.000$ & $-1.000$ & $-1.000$ & $0.911$ \\
OGLE-IV             & OGLE-GD-CEP-1376        & $  30.29682$ & $   3.17082$ & $  2341$ & $   76$ & $ 206$ & $   0.8713539$ & 1O & $-1.000$ & $-1.000$ & $-1.000$ & $-1.000$ & $ 9.441$ & $ 9.331$ & $ 9.339$ & $ 0.000$ & $0.594$ \\
OGLE-IV             & OGLE-GD-CEP-1371        & $  30.72821$ & $   4.15012$ & $ 21841$ & $  951$ & $ 164$ & $   3.3028235$ & 1O & $-1.000$ & $-1.000$ & $-1.000$ & $-1.000$ & $12.340$ & $12.323$ & $ 0.000$ & $ 0.000$ & $0.672$ \\
OGLE-IV             & OGLE-GD-CEP-1373        & $  30.75970$ & $   3.94171$ & $    -1$ & $    0$ & $  86$ & $   8.2963145$ & F  & $-1.000$ & $-1.000$ & $-1.000$ & $-1.000$ & $-1.000$ & $-1.000$ & $-1.000$ & $-1.000$ & $0.000$ \\
OGLE-IV             & OGLE-GD-CEP-1380        & $  31.62315$ & $   3.58712$ & $ 20839$ & $ 1055$ & $ 242$ & $   1.8913547$ & 1O & $-1.000$ & $-1.000$ & $-1.000$ & $-1.000$ & $13.091$ & $13.095$ & $ 0.000$ & $ 0.000$ & $0.750$ \\
OGLE-IV             & OGLE-GD-CEP-1375        & $  31.73567$ & $   4.17738$ & $ 16603$ & $  639$ & $ 133$ & $   3.0287829$ & 1O & $-1.000$ & $-1.000$ & $-1.000$ & $-1.000$ & $11.785$ & $11.778$ & $ 0.000$ & $ 0.000$ & $0.524$ \\
OGLE-IV             & OGLE-GD-CEP-1389        & $  31.89151$ & $   1.21029$ & $ 14163$ & $  366$ & $  91$ & $   4.7349844$ & 1O & $11.349$ & $11.260$ & $ 0.000$ & $ 0.000$ & $11.385$ & $11.546$ & $ 0.000$ & $ 0.000$ & $1.501$ \\
OGLE-IV             & OGLE-GD-CEP-1381        & $  32.13330$ & $   3.40401$ & $    -1$ & $    0$ & $ 140$ & $   2.5824883$ & 1O & $-1.000$ & $-1.000$ & $-1.000$ & $-1.000$ & $-1.000$ & $-1.000$ & $-1.000$ & $-1.000$ & $0.000$ \\
OGLE-IV             & OGLE-GD-CEP-1370        & $  32.21405$ & $   4.99390$ & $ 22272$ & $ 1265$ & $ 233$ & $   2.4750884$ & F  & $-1.000$ & $-1.000$ & $-1.000$ & $-1.000$ & $13.149$ & $13.171$ & $ 0.000$ & $ 0.000$ & $0.442$ \\
OGLE-IV             & OGLE-GD-CEP-1383        & $  32.61339$ & $   3.35610$ & $    -1$ & $    0$ & $ 193$ & $   2.0885693$ & F  & $-1.000$ & $-1.000$ & $-1.000$ & $-1.000$ & $-1.000$ & $-1.000$ & $-1.000$ & $-1.000$ & $0.000$ \\
OGLE-IV             & OGLE-GD-CEP-1398        & $  32.89903$ & $   0.06230$ & $  6140$ & $  189$ & $  83$ & $   6.5849492$ & F  & $ 9.157$ & $ 9.083$ & $ 8.881$ & $ 8.951$ & $-1.000$ & $-1.000$ & $-1.000$ & $-1.000$ & $0.573$ \\
OGLE-IV             & OGLE-GD-CEP-1387        & $  32.98170$ & $   2.48250$ & $ 20872$ & $ 1435$ & $ 172$ & $   3.6540989$ & F  & $-1.000$ & $-1.000$ & $-1.000$ & $-1.000$ & $12.634$ & $12.677$ & $ 0.000$ & $ 0.000$ & $0.818$ \\
OGLE-IV             & OGLE-GD-CEP-1391        & $  33.73762$ & $   1.39899$ & $ 13922$ & $  663$ & $ 111$ & $   4.9472037$ & F  & $11.622$ & $11.478$ & $ 0.000$ & $ 0.000$ & $11.265$ & $11.221$ & $ 0.000$ & $ 0.000$ & $0.939$ \\
OGLE-IV             & OGLE-GD-CEP-1395        & $  34.59883$ & $   1.41107$ & $ 11568$ & $  288$ & $  67$ & $  10.8186669$ & F  & $10.080$ & $ 9.962$ & $ 0.000$ & $ 0.000$ & $10.002$ & $ 9.932$ & $ 9.907$ & $ 0.000$ & $1.095$ \\
OGLE-IV             & OGLE-GD-CEP-1407        & $  34.85424$ & $  -0.05121$ & $  4091$ & $  108$ & $  51$ & $  17.0295145$ & F  & $ 7.424$ & $ 7.253$ & $ 7.136$ & $ 7.140$ & $ 7.234$ & $ 7.181$ & $ 0.000$ & $ 0.000$ & $1.522$ \\
OGLE-IV             & OGLE-GD-CEP-1405        & $  35.03510$ & $   0.33889$ & $ 19169$ & $  928$ & $  87$ & $   6.6459883$ & 1O & $12.143$ & $12.074$ & $12.175$ & $ 0.000$ & $-1.000$ & $-1.000$ & $-1.000$ & $-1.000$ & $2.772$ \\
OGLE-IV             & OGLE-GD-CEP-1409        & $  35.51655$ & $   0.03691$ & $  5155$ & $  144$ & $  51$ & $  16.3873766$ & F  & $ 7.896$ & $ 7.848$ & $ 7.679$ & $ 7.634$ & $ 7.738$ & $ 7.589$ & $ 0.000$ & $ 5.810$ & $1.456$ \\
OGLE-IV             & OGLE-GD-CEP-1388        & $  35.81171$ & $   3.91020$ & $    -1$ & $    0$ & $ 143$ & $   3.4726796$ & F  & $-1.000$ & $-1.000$ & $-1.000$ & $-1.000$ & $-1.000$ & $-1.000$ & $-1.000$ & $-1.000$ & $0.000$ \\
OGLE-IV             & OGLE-GD-CEP-1406        & $  35.89617$ & $   0.51846$ & $ 11064$ & $  761$ & $ 113$ & $   2.7327063$ & 1O & $11.707$ & $11.808$ & $12.069$ & $11.853$ & $-1.000$ & $-1.000$ & $-1.000$ & $-1.000$ & $1.939$ \\
OGLE-IV             & OGLE-GD-CEP-1390        & $  36.02872$ & $   3.26625$ & $ 19448$ & $  902$ & $ 230$ & $   1.8765092$ & 1O & $-1.000$ & $-1.000$ & $-1.000$ & $-1.000$ & $12.767$ & $12.696$ & $ 0.000$ & $ 0.000$ & $0.367$ \\
OGLE-IV             & OGLE-GD-CEP-1411        & $  36.74398$ & $   0.44648$ & $  2787$ & $   69$ & $  87$ & $   6.6266352$ & F  & $ 7.671$ & $ 7.529$ & $ 7.580$ & $ 7.544$ & $ 7.572$ & $ 7.437$ & $ 7.543$ & $ 0.000$ & $1.054$ \\
OGLE-IV             & OGLE-GD-CEP-1393        & $  36.91397$ & $   2.81453$ & $    -1$ & $    0$ & $ 169$ & $   2.6352329$ & F  & $-1.000$ & $-1.000$ & $-1.000$ & $-1.000$ & $-1.000$ & $-1.000$ & $-1.000$ & $-1.000$ & $0.000$ \\
OGLE-IV             & OGLE-GD-CEP-1403        & $  36.92415$ & $   1.35472$ & $ 13106$ & $  475$ & $ 104$ & $   5.5286975$ & F  & $11.347$ & $11.180$ & $ 0.000$ & $ 0.000$ & $-1.000$ & $-1.000$ & $-1.000$ & $-1.000$ & $1.163$ \\
\ldots              & \ldots                & \ldots       & \ldots       & \ldots   & \ldots  & \ldots & \ldots        &\ldots&\ldots   & \ldots   & \ldots   & \ldots   & \ldots   & \ldots   & \ldots   & \ldots   & \ldots  \\
\hline
\end{tabular}

\caption{The sample of newly detected classical Cepheids, sorted by Galactic
longitude. The columns are as follows:
1 -- The data source, 2~--~Cepheid ID, 3,4~--~Galactic coordinates, 
5,6 -- distance to the Cepheid and its error in pc, 7 -- Cepheid age in Myr,
8 -- Pulsation period in days, 9 -- Pulsation mode: fundamental (F), first-overtone (1O),
10--13 -- Spitzer photometry, 14--17~--~WISE photometry,
18 -- interstellar extinction in the Ks-band.
The full Table with additional 223 classical Cepheids is available from ftp://ogle.astrouw.edu.pl/ogle4/MILKY\_WAY\_3D\_MAP}
\end{sidewaystable}

\subsection{Classical Cepheid Distances}

Following the procedures described in detail in Skowron \etal (2019), we
used the mid-infrared (mid-IR) data from {\em Spitzer} (Benjamin \etal 2003;
Churchwell \etal 2009) and WISE (Wright \etal 2010; Mainzer \etal 2011) and
the mid-IR period-luminosity (PL) relations from Wang \etal (2018) to
calculate distances to the sample of 201 new classical Cepheids.
Extinction was estimated using the extinction maps of Bovy \etal (2016)
in the same way as in Skowron \etal (2019). Distances to classical Cepheids
are accurate to better than 5\%.

Fig.~3 shows the top view of the Milky Way with 2390 classical Cepheids 
that had mid-IR counterparts in the {\em Spitzer} and WISE catalogs. Objects
from Skowron \etal (2019) are marked with blue dots, while newly found
Cepheids with red dots.  The dotted line shows the
current extent of the OGLE survey (compare with Fig.~1B from Skowron
\etal 2019). The new classical Cepheids fill an almost empty region
(by far) that covers an area of the northern  Galactic warp, out to the
edge of the Milky Way disk.

\vspace{0.1cm}
\begin{figure*}[ht]
\centerline{\includegraphics[width=10cm]{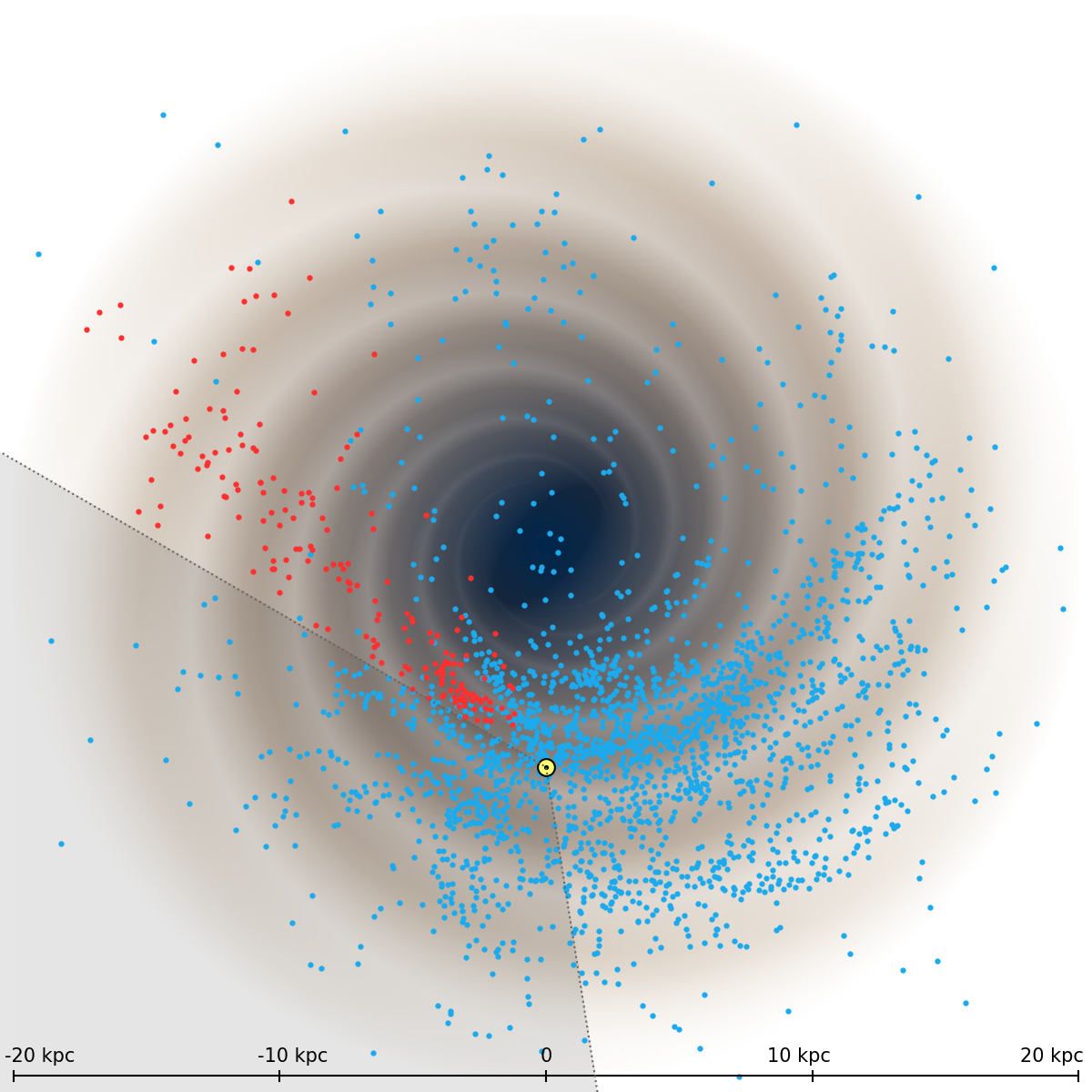}}
\FigCap{Top view of the Milky Way with all 2390 Cepheids with mid-IR
  counterparts. The sample from Skowron \etal (2019) is marked with blue
  dots, while the newly detected objects are marked with red dots.
  The background image represents a four-arm spiral galaxy model
  consistent with neutral hydrogen measurements in our Galaxy. The Sun
  is marked with a yellow disk, while the dotted lines show the
  angular extent of the extended OGLE footprint ($l<60\degg$ and
  $l>190\degg$).}
\vspace{0.3cm}
\label{fig:top}
\end{figure*}

\section{Results and Discussion}

\subsection{The Shape of the Warp}

We used the extended dataset of classical Cepheids to make an improved 3-D
model of the Galactic warp. Having a more complete coverage of the far side of
the disk than in Skowron \etal (2018) we attempted fitting the surface
in the form that includes two terms of the Fourier series in the
Galactocentric azimuth:
$$
z(R,\phi) =
\begin{cases}
z_0 & \quad \text{for~} R \leq R_d \\
z_0 + z_1 (R-R_d)^2 \sin(\phi-\phi_1) + z_2 (R-R_d)^2 \sin2(\phi-\phi_2) & \quad \text{for~} R>R_d
\end{cases}
\eqno(1)
$$
where $z$ is the vertical distance from the Galactic plane, $R$ is the distance
from the Galactic center, $\phi$ is the Galactocentric azimuth measured
counterclockwise from $l=0^{\circ}$, and $R_d$\,kpc is the radius at which
the disk starts warping; $z_0$, $z_1$, $z_2$, $\phi_1$, and $\phi_2$ describe
the shape of the surface.

The best-fitting parameters were found by minimizing the sum
$$
\chi^2 = \sum_i d_i^2 \exp(-0.5 (d_i/1\,\mathrm{kpc})^2), \eqno(2)
$$
where $d_i$ is an orthogonal distance between the surface and the $i$th star.
We multiplied the square of the distance by the exponential function to minimize
the impact of outliers (the median distance $d_i$ is 0.1\,kpc). We found that
the best-fit model is for: 
\begin{align*}
R_d &= 4.2258 \pm 0.1234 {\rm \;kpc}\\
z_0 &= 0.0447 \pm 0.0018 {\rm \;kpc}\\
z_1 &= 0.0089 \pm 0.0003 {\rm \;kpc^{-1}}\\
\phi_1 &= 158.3^{\circ} \pm 0.7^{\circ}\\
z_2 &= 0.0022 \pm 0.0001 {\rm \;kpc^{-1}} \\
\phi_2 &= -13.6^{\circ} \pm 2.1^{\circ}
\end{align*}

The uncertainties of these quantities were estimated with a Monte
Carlo simulation by drawing random heliocentric distances of Cepheids
from the Gaussian distributions. It is worth noting that the lines
of nodes of both terms of the Fourier series are very
correlated -- almost parallel.

\vspace{0.1cm}
\begin{figure*}[htb]
\centerline{\includegraphics[width=12cm]{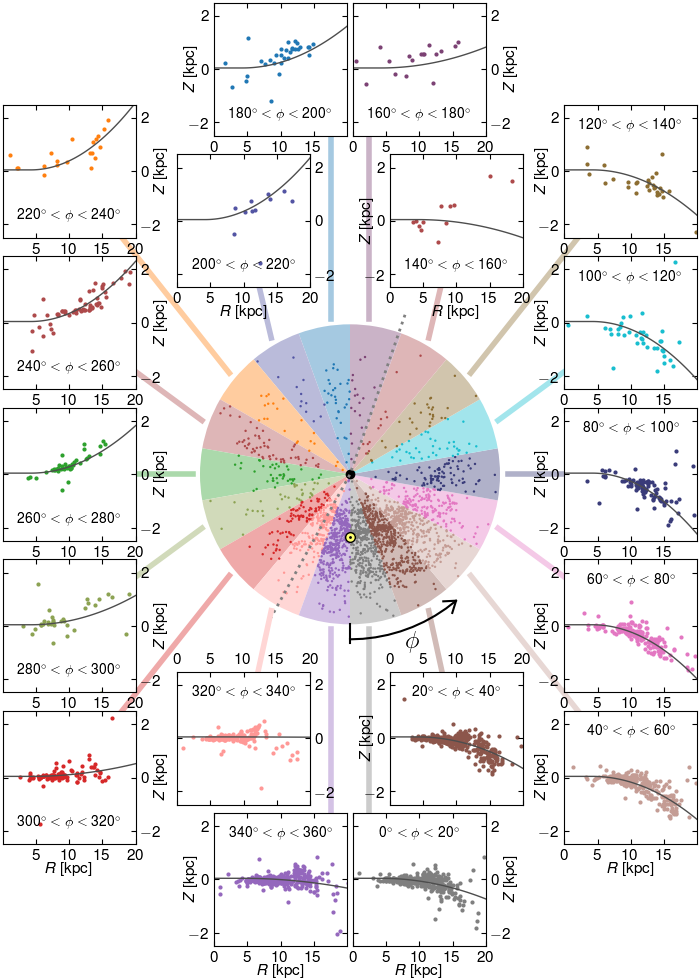}}
\FigCap{The new model of the Galactic warp. The disk is divided into 18
  regions in the Galactocentric polar coordinate system, marked with
  different colors (figure center). The Sun is marked with a yellow dot
  while the Galactic center with a black dot. The azimuth $\phi=0^{\circ}$
  is pointing toward the Sun. The dotted line divides parts of the
  Milky Way disk model warped south and north.
  The vertical distribution of Cepheids  within each region is shown in
  corresponding panels with matching colors, where the Galactocentric
  distance $R$ is plotted against distance from the Galactic plane $Z$.
  Black lines show the intersection of the median model surface within
  a given region.}
\vspace{0.3cm}
\label{fig:warp}
\end{figure*}

Fig.~4 presents the new model of the Galactic warp. The disk in the center
of the figure shows the top view of the Galaxy, with individual Cepheids marked
with colored dots. The disk is divided into 18 regions in the Galactocentric
polar coordinate system, each region being $20^{\circ}$ wide in azimuth and
marked with a different color.
The vertical distribution of Cepheids  within each region is shown in
corresponding side panels with matching colors, where the Galactocentric
distance $R$ is plotted against distance from the Galactic plane $Z$.
Black lines show the intersection of the median model surface within
a given region.

Fig.~4 shows a good agreement between the new model and the observed
distribution of Cepheids. A simpler, three parameter model (where $z_2$ is set
to zero) also provides a reasonable fit, but it cannot explain the global
shape of the disk, especially in the range $90^{\circ}<\phi<240^{\circ}$.
Note that instead of using a fixed $R_d=8$\,kpc (Skowron \etal 2019), we leave
it as a free parameter to obtain a better fit. This is now possible because
the number of objects in the azimuth range $90^{\circ}<\phi<270^{\circ}$ is
sufficient.

Fig.~4 also indicates that our Cepheid sample populates all regions of the
Galactic disk very well, thus presenting a complete picture of the
distribution of young stars in the Milky Way, even though there are a few
less populated sectors on the far side of the Galactic center, \ie
$140^{\circ}<\phi<160^{\circ}$, $160^{\circ}<\phi<180^{\circ}$, and
$200^{\circ}<\phi<220^{\circ}$. Unfortunately, large interstellar extinction
in these directions makes Cepheid detection very difficult in the optical
range (see Fig.~1). In principle one could complement this sample with
Cepheids detected in the infrared (IR) range, where the extinction is much
lower. However, Cepheids are much more difficult to identify in the IR
because their light curves are less unique and often resemble other types
of variable stars. The IR sample of Cepheids based on the VISTA Variables
in The Via Lactea (VVV) survey data  was recently released by D\'ek\'any
\etal (2019). We verified those objects that had counterparts in the OGLE data
($I<19$~mag; over $300$ stars) and found that only about 55\% were in fact
genuine Cepheids; the remaining ones ($\sim40\%$) were eclipsing binaries
or spotted stars. For this reason we decided not to include the Cepheid sample
of D\'ek\'any \etal (2019) in our analysis.

\vspace{0.1cm}
\begin{figure*}[ht]
\centerline{\includegraphics[width=12cm]{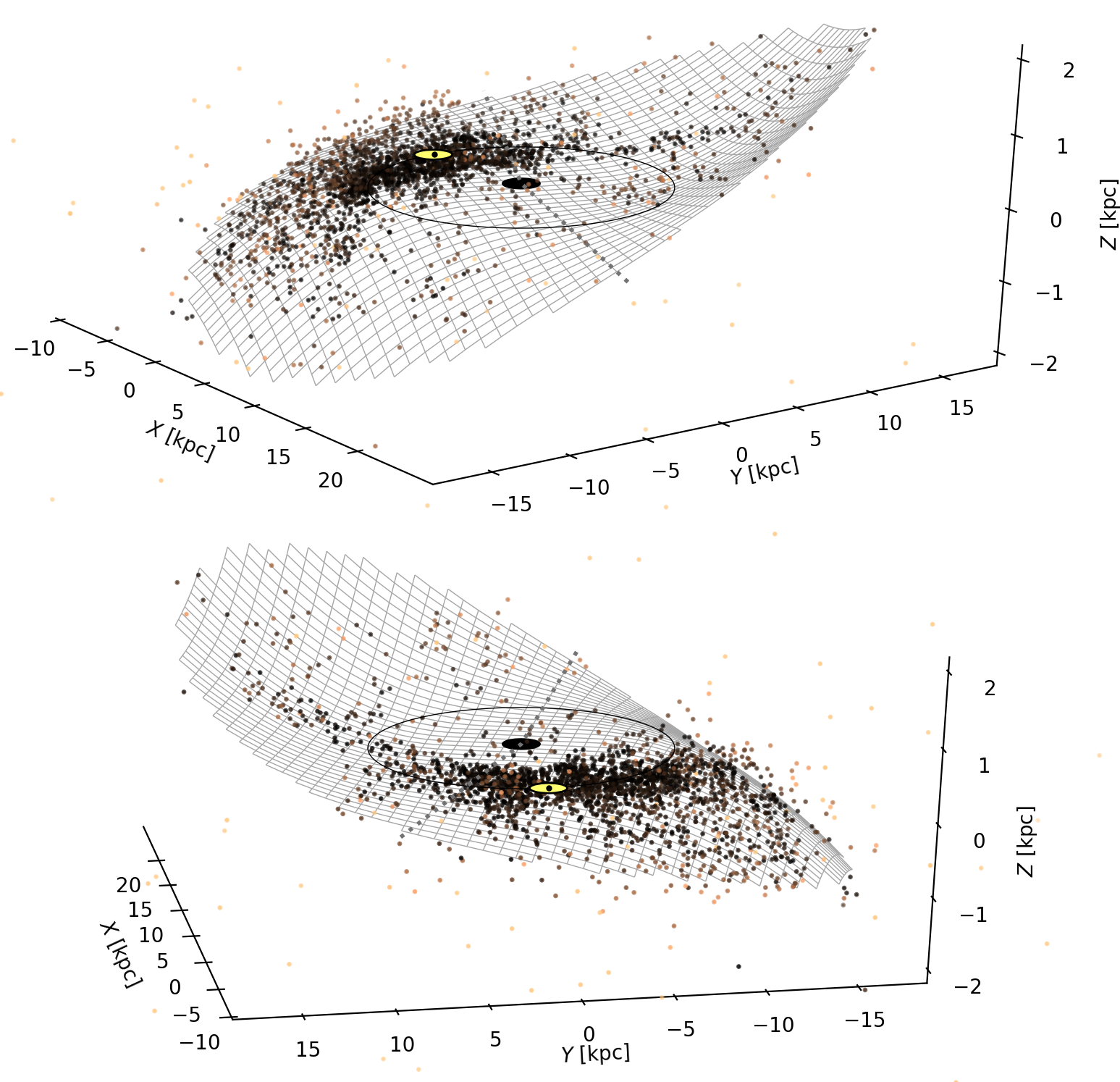}}
\FigCap{Three dimensional view of the Milky Way in young stellar
  population. Dots mark positions of classical Cepheids from our
  sample. The grid represents our new model of the Galactic disk
  (Eq.1). The viewing angles are $325^{\circ}$ (top) and $170^{\circ}$
  (bottom), at an inclination angle $30^{\circ}$. The Sun is marked
  with a yellow dot, while the solid line represents the solar circle.
  The dashed lines mark the position of the lines of nodes.}
\vspace{0.3cm}
\label{fig:3d}
\end{figure*}

Fig.~5 presents the best 3-D five parameter model of the Milky Way
as the gray grid, with colored dots marking the positions of
individual Cepheids from our sample. The viewing angles are $325^{\circ}$
(top) and $170^{\circ}$ (bottom), at an inclination angle $30^{\circ}$.

\vspace{0.1cm}
\begin{figure*}[ht]
\centerline{\includegraphics[width=12.5cm]{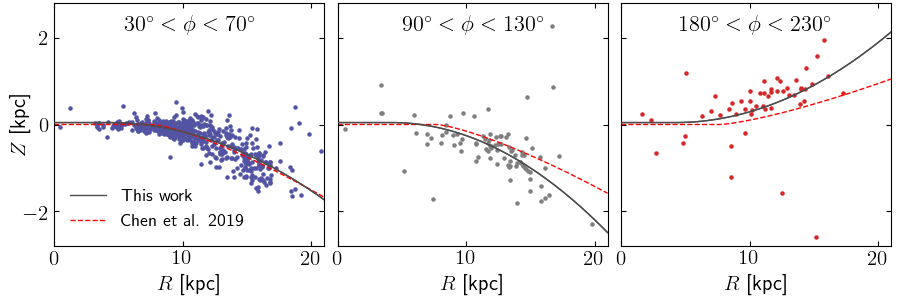}}
\FigCap{The comparison of the Milky Way models from Chen \etal (2019) and
  this work for three selected regions of the Milky Way disk. The power law
  model of Chen \etal (2019) is marked with a red dashed line, while the 
  model from this work with a gray solid line.}
\vspace{0.3cm}
\label{fig:chen}
\end{figure*}

The comparison of the distribution of our Cepheid sample and the fitted model
with the power law model of the Galactic warp presented by
Chen \etal (2019) is shown in Fig.~6 for three selected non-adjacent sectors.
We see that the agreement between models is reasonable in the range
$30^{\circ}<\phi<70^{\circ}$ (left panel), which is the region well populated with
Cepheids in both samples. However, on the far side of the Milky Way (middle and
right panels), the model of Chen \etal (2019) differs significantly from our
model. This is not surprising because the sample of Cepheids used by Chen \etal
(2019) to derive their models was much smaller and located mostly in the second
and third Galactic quadrants, thus not covering the region of the northern
Galactic warp.

Finally, the left panel of Fig.~7 shows the vertical displacement of Cepheids
from the Galactic plane. Interestingly, the center of the flat part of the
Galactic disk ($|Z| \simeq 0$) does not coincide with the Galactic center
(black dot). If we model the warp with a shift of the warp center toward the
Sun as a new free parameter, we find that the second term in the Fourier series
(Eq.~1) is no longer necessary, while the warp center moves $2.7$\,kpc away from
the Galactic center. 
The remaining parameters of the fit are:
$R_d = 4.47{\rm \;kpc}$,
$z_0 = -0.001{\rm \;kpc}$,
$z_1 = 0.010{\rm \;kpc^{-1}}$,
$\phi_1 = 159^{\circ}$.
Allowing the fit to move in the Y direction does not produce a significantly
better fit. The fact that the shift of the warp center correlates with the
line connecting the Sun with the Galactic center suggests that the shift may
be caused by the incompleteness of the Cepheid sample in the highly extincted
area around the Galactic center.
We therefore repeated the fit after removing Cepheids that fall into the
regions where their detection probability by the OGLE survey is less than
80\% (gray and black area in Fig.~1). The results of the fit change only
slightly, with the shift of $1.8$\,kpc, $R_d = 5.0{\rm \;kpc}$, and
$\phi_1 = 163^{\circ}$, meaning that this is not solely a selection effect.
The asymmetry of the S-shaped warps seems to be fairly common in simulations
of warped disks caused by interactions with a satellite galaxy on a highly
elongated orbit (Kim \etal 2014).

\subsection{The Age Dependency of the Warp}

The age dependence of the Milky Way warp, both in position and kinematics,
has been investigated by Am\^ores \etal (2017) with the use of the 2MASS data
and a population synthesis model. They found that there is a clear dependence
of the thin disk scale length, warp and flare shapes with age, such that the
thin disk scale length for the youngest population is twice the scale length
for the oldest stars. The warping amplitude was found to be larger in the
area of the northern warp for all population ages (0.5--6.0~Gyr).

Romero-G\'{o}mez \etal (2019) used the astrometric information from the
{\em Gaia} Data Release 2 (DR2; {\em Gaia} Collaboration \etal 2016, 2018)
to analyze the distribution of a sample of young
stars (mainly OB type) and a sample of red giant branch (RGB) stars. They
showed that there is the age dependency of the Galactic warp, both in position
and kinematics, such that the young OB stars are less warped than the RGB stars
(the height of 0.2 kpc vs. 1.0 kpc at a Galactocentric distance of 14 kpc,
respectively).
Also the onset radius of the warp is different for the two groups, \ie 12--13
kpc for the young stars and 10--11~kpc for the RGB stars. This suggests
that the warping of the disk is more pronounced in the older populations.

Classical Cepheids are young objects, thus their distribution should be similar
to that of OB stars. However, the findings of Romero-G\'{o}mez \etal (2019)
are in contradiction to our results. We found that the onset radius of the warp
for Cepheids ($\sim4$~kpc) is much lower than for both samples of
Romero-G\'{o}mez \etal (2019), and the median height of Cepheids at
a Galactocentric distance of 14 kpc is about 1~kpc, which is similar to the RGB
sample, and clearly higher than for the OB sample. 

Our data show that the northern part of the Galactic warp has a larger amplitude 
than the southern by about 200~pc at 14~kpc, while the RGB data in
Romero-G\'{o}mez \etal (2019) showed the opposite. On the other hand, the
analysis by Am\^ores \etal (2017) confirms the asymmetry seen with classical
Cepheids  -- with the northern warp being more pronounced than the southern,
although the degree of the asymmetry is different
(see Table~1 in Romero-G\'{o}mez \etal 2019).

It is worth noting, however, that the discussed Galactic disk models require
various assumptions that may influence the final result. For example, Poggio
\etal (2018) and Romero-G\'{o}mez \etal (2019) compute distances using a
Bayesian estimator (and {\em Gaia} parallaxes) whereas Am\^ores \etal (2017)
use 2MASS star counts and color--magnitude diagrams to constrain their
analytical model of the warp.
On the contrary, Galactic disk models based on Cepheids are free of any 
assumptions and rely solely on direct and accurate distances to individual
stars.

\subsection{The Kinematics of the Warp}

In addition to the spatial 3-D distribution of classical Cepheids, kinematical
data may provide useful information about the warp. Previously, Poggio \etal
(2018) used the {\em Gaia} DR2 data to find a gradient of $5-6$\,km/s in the
vertical velocities of upper main sequence stars and giants located from 8 to
14\,kpc in Galactic radius. A similar kinematical evidence of the warp was
detected by Romero-G\'omez \etal (2019).

The warping is most likely to manifest in a component of the velocity vector
perpendicular to the Galactic plane:
$$
W = v_b \cos{b} + v_r \sin{b} + W_{\odot},
$$
where $v_b = 4.74 \mu_b D$, $\mu_b$ is the proper motion in Galactic latitude,
$D$ -- distance, $v_r$ -- radial velocity, and $W_{\odot}=7.3$\,km/s is the
vertical component of the Solar velocity relative to the local standard of rest
(LSR; Sch{\"o}nrich \etal 2010).

For the majority of detected Cepheids, we lack information about their radial
velocities. Fortunately, the radial component can be estimated by assuming
that random velocities of Cepheids are much smaller than their total velocity
around the Galactic center. We assume the linear Galactic rotation velocity
curve of Mr\'oz \etal (2019) and convert it to the expected radial velocity
by using the formalism of Reid \etal (2009). By comparing the estimated radial
velocity and actual {\em Gaia} measurements for 866 objects, we find a
dispersion of $18.6~{\rm km/s}$, which translates into the vertical velocity
spread of $18.6~{\rm km/s} \times \sin{b} \leq 2-3~{\rm km/s}$.

\vspace{0.1cm}
\begin{figure*}[ht]
\centerline{\includegraphics[width=13cm]{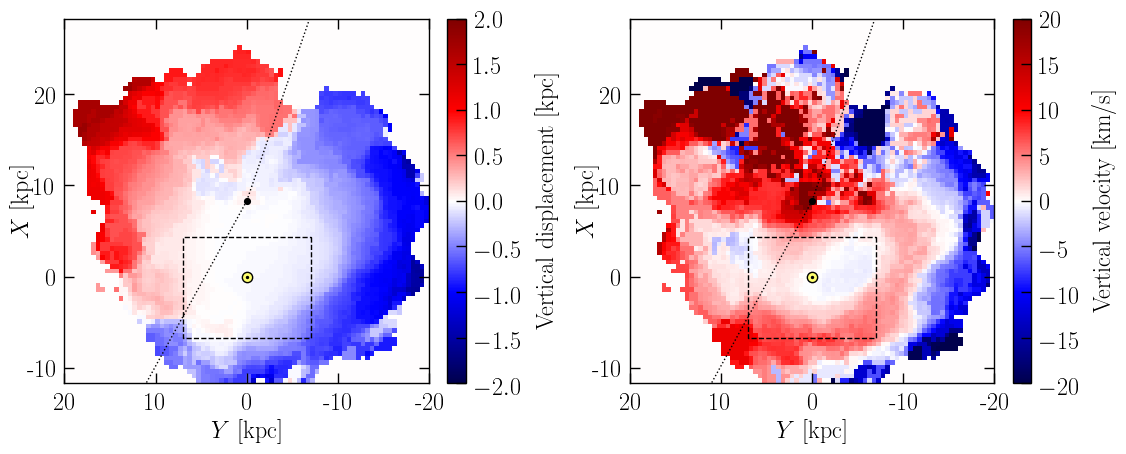}}
\FigCap{The vertical distance $Z$ from the Galactic plane (left panel) showing
  the extent and amplitude of the Galactic warp. The vertical velocity
  distribution (right panel) is smoothed by calculating a median velocity
  within 3\,kpc of $0.5\times 0.5$\,kpc wide bins. The dotted lines represent
  the lines of nodes, the dashed line box shows an area investigated by Poggio
  \etal (2018). The Sun is marked with a yellow dot while the Galactic center
  with a black dot.}
\vspace{0.3cm}
\label{fig:chen}
\end{figure*}

Then, we smooth the vertical velocity distribution by calculating a median
velocity within 3\,kpc of $0.5\times 0.5$\,kpc wide bins (right panel of
Fig.~7). The resulting map presents a complicated pattern of vertical
velocities with a strong gradient in a direction perpendicular to the line of
nodes of the warp. Cepheids located in the warp below the Galactic plane show
large negative vertical velocity ($-20 \div -10$\,km/s) whereas those located
in the northern warp are moving toward the north Galactic pole (with positive
vertical velocities of $10 \div 20$\,km/s). Cepheids can be detected up to the
edge of the Galactic disk whereas stars analyzed by Poggio \etal (2018) are
located much closer to the Sun (approximately, $-7 <X< 5$\,kpc, $|Y|< 7$\,kpc).
In the overlapping area (the dashed line box in Fig.~7), our findings are
similar to those of Poggio \etal (2018; see their Fig.~3CD). However, our map
reveals systematic vertical motions on a much larger scale.

The vertical velocity change with the distance from the Galactic center
(\ie vertical velocity waves), could be induced by the Sagittarius dwarf
galaxy (G\'omez \etal 2013) together with the vertical density waves in the Milky
Way disk. The vertical velocity distribution of our Cepheid sample looks
very similar, both in space and amplitude, to theoretical models (Fig.~6E of
G\'omez \etal 2013) produced by a simulation of the Sagittarius impact
into the plane of the Milky Way. On the other hand, we do not observe
the vertical density waves that should be phase-shifted with respect to the
velocity waves.

\section{Summary}

We have constructed the 3-D map of the Milky Way based on classical Cepheids.
The map is an update of the one presented in Skowron \etal (2019),
supplemented with about 200 newly detected objects from the extension of the
OGLE footprint in the Galactic disk, observed during the 2018 and 2019 seasons
(Udalski \etal 2018).
New Cepheids are predominantly distributed in the first Galactic quadrant
-- the region that was sparsely populated with Cepheids so far and that is
directly opposite the region where we observe the largest bending of the
Galactic disk in the southern direction (the third Galactic quadrant).

Direct distances were determined based on the mid-IR data from {\em Spitzer}
and WISE and are accurate to better than 5\%.
We modeled the distribution of Cepheids with an analytic formula including
two terms of the Fourier series in Galactic azimuth (Eq.1). We found that this
model fits the entire Galactic disk very well (Fig.~4).
The northern warp is very prominent and its amplitude is slightly higher than
in the southern part, indicating the asymmetry of the warp.

We use \textit{Gaia} astrometric data together with distances to Cepheids
from our sample and the Galactic rotation curve to construct a map of the
vertical component of the velocity vector for all Cepheids in the Milky Way
disk. We find that Cepheids located in the northern warp have large positive
vertical velocity (toward the north Galactic pole), while those in the
southern warp -- negative vertical velocity (toward the south Galactic pole).
The vertical velocity amplitudes are of the order of 10--20 km/s.

Because the sample of Cepheids analyzed in this work populates almost an entire
Galactic disk, there is not much room for further improvements of the model
of the Galactic warp. Additional Galactic classical Cepheids can still be found
in the regions of the northern sky unavailable to the OGLE survey, and with the
use of infrared observations. These regions are, however, covering
mostly the flat parts of the disk close to the lines of nodes of the disk
model.


\Acknow{The OGLE project has received funding from the National Science Center
(NCN) grant MAESTRO 2014/14/A/ST9/00121 to AU. This work has been supported in
part by the NCN grant MAESTRO no. 2016/22/A/ST9/00009 to IS.
This work has made use of data from the European Space Agency (ESA) mission
{\it Gaia} ({\em https://www.cosmos.esa.int/gaia}), processed by the {\it Gaia}
Data Processing and Analysis Consortium (DPAC,\\
{\em https://www.cosmos.esa.int/web/gaia/dpac/consortium}). Funding for the DPAC
has been provided by national institutions, in particular the institutions
participating in the {\it Gaia} Multilateral Agreement.
}



\begin{references}
\refitem{Am\^ores, E.~B., Robin, A.~C., Reyl\'e, C.}{2017}{\AA}{602}{A67}
\refitem{Benjamin, R.~A., \etal}{2003}{\PASP}{115}{953}		
\refitem{Berdnikov, L.~N.}{1987}{Pis'ma Astron. Zh.}{13}{45}
\refitem{Bovy, J., Rix, H.-W., Green, G.~M., Schlafly, E.~F., Finkbeiner, D.~P.}{2016}{\ApJ}{818}{130}
\refitem{Burke, B.~ F.}{1957}{\AJ}{62}{90}
\refitem{Chen, X., Wang, S., Deng, L., de Grijs, R., Yang, M.}{2018}{\ApJS}{237}{28}
\refitem{Chen, X., Wang, S., Deng, L., de Grijs, R., Liu, C., Tian, H.}{2019}{Nature Astronomy}{3}{320}
\refitem{Churchwell, E., \etal}{2009}{\PASP}{121}{213}		
\refitem{D\'ek\'any, I., Hajdu, G., Grebel, E.~K., Catelan, M.}{2019}{\ApJ}{883}{58}
\refitem{Drimmel, R., Spergel, D.~N.}{2001}{\ApJ}{556}{181}
\refitem{Gaia Collaboration, Prusti, T., \etal}{2016}{\AA}{595}{1}	
\refitem{Gaia Collaboration, Brown, A.~G.~A., \etal}{2018}{\AA}{616}{1}	
\refitem{Gieren, W.~P., Fouqu\'e, P., G\'omez, M.}{1998}{\ApJ}{496}{17}
\refitem{G\'omez, F.~A., Minchev, I., O'Shea, B.~W., Beers, T.~C., Bullock, J.~S., Purcell, C.~W.}{2013}{\MNRAS}{429}{159}
\refitem{Holl, B., \etal}{2018}{\AA}{618}{30}			
\refitem{Kim, J.~H., Peirani, S., Kim, S., Ann, H.~B., An, S.-H., Yoon, S.-J.}{2014}{\ApJ}{789}{90}
\refitem{Levine, E.~S., Blitz, L., Heiles, C.}{2006}{Science}{312}{1773}
\refitem{L\'opez-Corredoira, M.}{2019}{arXiv:}{1909.09815}{.}
\refitem{Mainzer, A., \etal}{2011}{\ApJ}{731}{53}		
\refitem{Marshall, D.~J., Robin, A.~C., Reyl\'e, C., Schultheis, M., Picaud, S.}{2006}{\AA}{453}{635}
\refitem{Masci, F.~J., \etal}{2019}{\PASP}{131}{995}		
\refitem{Mr\'oz, P., \etal}{2019}{\ApJ}{870}{10}
\refitem{Nakanishi, H., Sofue, Y.}{2015}{PASJ}{58}{847}
\refitem{Poggio, E., \etal}{2018}{\MNRAS}{481}{L21}
\refitem{Pojma\'nski. G.}{2002}{\Acta}{52}{397}
\refitem{Reid, M.~J., Menten, K.~M., Zheng, X.~W., \etal}{2009}{\ApJ}{700}{137}
\refitem{Reyl\'e, C., Marshall, D.~J., Robin, A.~C., Schultheis, M.}{2009}{\AA}{495}{819}
\refitem{Romero-G\'omez, M., Mateu, C., Aguilar, L., Figueras, F., Castro-Ginard, A.}{2019}{\AA}{627}{A150}
\refitem{Sanchez-Saavedra, M.~L., Battaner, E., Florido, E.}{1990}{\MNRAS}{246}{458}
\refitem{Sch{\"o}nrich, R., Binney, J., Dehnen, W.}{2010}{\MNRAS}{403}{1829}
\refitem{Skowron, D.~M., \etal}{2019}{Science}{365}{478}
\refitem{Smart, R. L., Drimmel, R., Lattanzi, M.~G., Binney, J.~J.}{1998}{Nature}{392}{471}
\refitem{Udalski, A., Szyma\'nski, M.~K., Szyma\'nski, G.}{2015}{\Acta}{65}{1}
\refitem{Udalski, A., \etal}{2018}{\Acta}{68}{315}		
\refitem{Wang, S., Chen, X., de Grijs, R., Deng, L.}{2018}{\ApJ}{852}{78}
\refitem{Westerhout, G.}{1957}{Bull. Astron. Inst. Neth.}{13}{201}
\refitem{Wright, E.~L., \etal}{2010}{\AJ}{140}{1868}		
\end{references}
\end{document}